\documentclass[aps,floatfix,showpacs,10|pt,superscriptaddress]{revtex4-2}
\usepackage{graphicx}
\usepackage[compat=1.1.0]{tikz-feynman}
\usepackage{amsmath, amsfonts, amssymb, bm}
\usepackage[vmargin=2.5cm, hmargin=3.0cm]{geometry}
\usepackage{slashed}
\usepackage{calrsfs}
\usepackage{hyperref}
\usepackage{cleveref}
\usepackage{appendix}
\usepackage{soul}

\def \[#1\]{\begin{equation}\begin{split}
			#1
	\end{split}\end{equation}
}

\newcommand{\inv}[1]{\frac{1}{#1}}
\newcommand{\Ogrande}{\mathcal{O}}	

\newcommand{\barr}[1]{\overset{}{argument}  }

\newcommand{\braket}[2]{\langle #1|#2\rangle}

\newcommand{\pder}[2]{\frac{\partial #1}{\partial #2}}

\newcommand{\abs}[1]{\left|#1\right|}

\DeclareMathAlphabet{\pazocal}{OMS}{zplm}{m}{n}

\tikzfeynmanset{
	graviton/.style={
		decoration={zigzag,amplitude = 0.5 mm,segment length=1.5 mm,pre,pre length=2pt, post,post length=1pt},decorate, double}
}
\tikzfeynmanset{
	gravitonbg/.style={
		decoration={ zigzag,amplitude = 0.5 mm,segment length=1.5mm,pre,pre length=2pt, post,post length=1pt},decorate,double ,very thick}
}
\tikzfeynmanset{
	photonbg/.style={
		decoration={ snake,amplitude = 0.5 mm,segment length=1.5mm,pre,pre length=2pt, post,post length=1pt},decorate,
		very  thick }
}
\tikzfeynmanset{doublefermion/.style={
		/tikz/double,
		/tikz/decoration={name=none},
		/tikz/postaction={
			/tikzfeynman/with arrow=0.5,
		}
	}
}
\tikzfeynmanset{doublescalar/.style={
		/tikz/double,
		/tikz/dashed,
		decoration={ name = none},
		/tikz/postaction={
			/tikzfeynman/with arrow=0.5,
		}
	}
}

\definecolor{auburn}{rgb}{0.43, 0.21, 0.1}
\definecolor{burgundy}{rgb}{0.5, 0.0, 0.13}
\definecolor{burntorange}{rgb}{0.8, 0.33, 0.0}
\definecolor{amethyst}{rgb}{0.6, 0.4, 0.8}
\definecolor{darkcerulean}{rgb}{0.03, 0.27, 0.49}
\definecolor{applegreen}{rgb}{0.55, 0.71, 0.0}
\definecolor{chocolate}{rgb}{0.82, 0.41, 0.12}
\definecolor{darkgreen}{rgb}{0.0, 0.5, 0.0}
\definecolor{violet}{rgb}{0.54, 0.17, 0.89}
\definecolor{azzurro}{rgb}{0.0, 0.44, 1.0}

\begin{abstract}
The classical dynamics and the construction of quantum states in a plane wave curved spacetime are examined, paying particular attention to the similarities with the case of an electromagnetic plane wave in flat spacetime. A natural map connecting the dynamics of a particle in the Rosen metric and the motion of a charged particle in an electromagnetic plane wave is unveiled. We then discuss how this map can be translated into the quantum description by exploiting the large number of underlying symmetries. We examine the complete analogy between Volkov solutions and fermion states in the Rosen chart and properly extend this to massive vector bosons. We finally report the squared $S$-matrix element of Compton scattering in a sandwich plane wave spacetime in the form of a two-dimensional integral.   

\end{abstract}
\begin{document}
	\author{G. Audagnotto}
	\email{audag@mpi-hd.mpg.de}
	\affiliation{Max Planck Institute for Nuclear Physics, Saupfercheckweg 1, D-69117 Heidelberg, Germany}
	\author{A. Di Piazza}
	\email{a.dipiazza@rochester.edu}
 \affiliation{Max Planck Institute for Nuclear Physics, Saupfercheckweg 1, D-69117 Heidelberg, Germany}
\affiliation{Department of Physics and Astronomy, University of Rochester, Rochester, New York 14627, USA}
\affiliation{Laboratory for Laser Energetics, University of Rochester, Rochester, New York 14623, USA}
	
%
 
 \title{ Dynamics, quantum states and Compton scattering in nonlinear gravitational waves }	
	\maketitle

    \section{Introduction}
	Gravitational waves that reach the Earth are generally weak. To have an idea, the typical fractional deformation coming from astrophysical sources is of the order of $ \sim 10^{-21}$ or less. This means that Earth-based detectors like LIGO, Virgo or KAGRA, which have sizes of the order of $10^3 \text{m}$, need the incredible displacement sensitivity of $\sim 10^{-18} \text{m}$ in order to detect a gravitational perturbation \cite{Bailes:2021,Flanagan:2005yc, LIGOScientific_2016}. This is the reason why gravitational waves are usually treated in the weak field approximation \cite{Weinberg_1972}, in which the spacetime metric $g_{\mu \nu}$ is approximated as the flat metric $\eta_{\mu\nu}$ plus a small correction $\kappa h_{\mu\nu}$ and a first-order treatment of the latter is in most cases enough for any measurable prediction. 
    Despite the weakness of the perturbation amplitudes typically measured on Earth, nonlinear gravitational plane waves, as exact solutions of Einstein's equations, can become an interesting subject for different reasons. One motivation comes from the fact that higher order corrections can grow substantially with the distance between the source and the observer, making nonlinear effects eventually not negligible. This has been pointed out, for example, in Refs. \cite{Harte:2015ila,Harte:2012jg}. 
	This is due to the fact that the dynamics can be expressed in terms of a matrix $e_{ij}$ satisfying the harmonic equation $\ddot{e}_{ij} = H_{ik}e^k_{\-\ j}$, where the profile $H_{ij}$ encodes the spacetime curvature as a function of the wave phase (see Sec. \ref{sec. Brinkmann and Rosen}). For small amplitudes one can identify $2 H_{ij} =\kappa \ddot{h}_{ij}$. However, this does not necessarily imply that only the first-order term in $e_{ij}$ has to be considered, in fact the second-order correction involves an integration of $(\dot{h}^i_{\-\ i})^2$ and this generally grows with the phase length of the wave \cite{Harte:2015ila}. These large-scale effects could in principle affect pulsar timing measurements \cite{Hobbs:2009yy, Harte:2015ila}, which very recently evidenced the presence of low-frequency background gravitational waves \cite{NANOGrav:2023gor}.
    Nonlinear effects are interesting also because they can be of a different nature than the linear ones, position and velocity memory effects provide an example that attracted a lot of attention in the last years as well \cite{Braginsky:1985vlg, Harte:2015ila,Harte:2012jg,  Grishchuk_1977, Zhang:2017rno}.
    Another reason for which exact gravitational plane waves are worth to be studied comes from the so-called Penrose limit \cite{Penrose1976, Blau:2002mw}. Penrose proved that ``any spacetime has a plane wave as a limit'', namely the spacetime in a small region around a null geodesic assumes a plane wave form. This is the gravitational analog of a well known fact in electromagnetism, indeed an observer moving at ultrarelativistic velocities perceives an arbitrary electromagnetic field approximately as a plane wave and the faster the observer moves the more accurate is the similarity \cite{Jackson_b_1975}. The generality of this property suggests that despite being idealizations, exact plane waves can provide an interesting scenario for studying limiting cases.
	
    The paper is organized as follows, in Sec. \ref{Plane_Waves} we discuss how plane waves in flat spacetime can be defined through their symmetries and the role that these play in the dynamics. Section \ref{sec. Brinkmann and Rosen} is devoted to the description of Brinkmann and Rosen charts, which provide two somehow complementary ways of describing a nonlinear gravitational wave. In Sec. \ref{sec. QF} we examine the construction of quantum states in a plane wave spacetime for scalar, spinor, and vector particles, underlining the similarities with the flat spacetime case deriving from shared symmetries and extending the results found in Ref. \cite{Adamo_2019}. Finally, in Sec. \ref{sec. Compton} we provide the full spin and polarization summed squared $S$-matrix element for Compton scattering in a nonlinear gravitational wave background.

	\subsection{Conventions}
	The signature of the metric chosen in this work is -2 and we assume $\hbar = c=1$, $\kappa = \sqrt{32 \pi G}$. Referring to the Misner-Thorne-Wheeler systematization \cite{Misner:1973prb} we adopt the convention $[-,-,-]$.
	Plane waves propagating along $\bm{z}$ depend on the variable $t-z$ only. For this reason we will choose to work with light-cone coordinates
	 $\{{x^+, x^i, x^-} \}$ defined as $x^- = t-z \equiv \phi$, $x^+ = \frac{1}{2}(t+z)$, $x^i = \{x,y\}$. The Latin indices will be used throughout this paper referring to the two transverse coordinates. In general, we can introduce the four quantities $\{{n^\mu, \delta_i^\mu, \tilde{n}^\mu} \}$ with $n^\mu = (1,0,0,1)$, $\tilde{n}^\mu = \frac{1}{2}(1,0,0,-1)$ such that any vector $v^\mu $ can be decomposed as $v^\mu = v^- \tilde{n}^\mu + v^+ n^\mu + v^i \delta_i^\mu$ with $v^- = n \cdot v$, and $v^+ = \tilde{n} \cdot v$. Letters belonging to the first half of the Greek alphabet $\alpha$, $\beta$,... will refer to the flat spacetime metric unless otherwise stated, whereas $\mu$, $\nu$,... will be used as curved spacetime indices. The dot symbol will only be used for flat spacetime products such that $v\cdot w = v^\alpha w^\beta \eta_{\alpha \beta}$ but $ v^\mu w^\nu g_{\mu \nu} \neq v\cdot w $. 
	  Overdots $\dot{f}$ will represent derivatives with respect to plane wave phase $\phi$, as already done in the introduction. Quantities projected
over the Rosen metric's vierbein $e_{\alpha \mu}$ are indicated with an overbar $\bar{v}^\alpha = e^\alpha_{\-\ \mu }v^\mu$, consequently $\bar{\gamma}^\alpha$ are the usual Dirac matrices.
The slashed notation will only be used for contractions between
flat spacetime indices and Dirac matrices $\slashed{v} = \bar{\gamma}^\alpha v_\alpha$.
Symmetrization and antisymmetrization are defined as 
	 $
    T_{(\mu_1 .. \mu_n )}  = \inv{n!} \sum_{\sigma \in S_n} T_{\mu_{\sigma(1)}... \mu_{\sigma(n)}}$
  and $
	 T_{[\mu_1 .. \mu_n ]}  = \inv{n!} \sum_{\sigma \in S_n} \text{sgn}(\sigma) T_{\mu_{\sigma(1)}... \mu_{\sigma(n)	}}
	 $, respectively, where $S_n$ indicates the permutations of $n$ indices and $\text{sgn}$ is the sign function.

	\section{Plane waves and their symmetries}
 \label{Plane_Waves}
	What is a plane wave? We are used to think about a plane wave $\Phi(\phi)$ as a field that depends only on the combination $\phi = n\cdot x = t-z$ of spacetime coordinates, with $\bm{z}$ identifying the propagation direction of the plane wave itself. This intuitive idea unveils three symmetries, namely translations along all the spacetime coordinates except $\phi$, i.e. $x^+=(t+z)/2$ and $x^i$, where $i=1,2$ refers to the coordinates transverse to $\bm{z}$. A more careful investigation reveals that plane waves feature two more symmetries $\chi_i$ belonging to the Lorentz group. These Killing vector fields correspond to the infinitesimal Lorentz transformations $\Lambda^{\alpha}{}_{\beta}=\delta^{\alpha}{}_{\beta}+\omega^{\alpha}{}_{\beta}$ leaving $n^\alpha$ unchanged, namely
	$
	\omega_{\alpha\beta}n^\beta = 0
	$. This equation defines the massless Wigner little group associated to the null vector $n^\alpha$ (see e.g. \cite{Wigner:1939cj,Bargmann:1948ck,Weinberg:1995mt}). The Lorentz coefficients satisfying this equation are a linear combination of the two antisymmetric tensors $\mathcal{F}_i^{\alpha\beta} = n^\alpha \delta_i^\beta -n^\beta \delta_i^\alpha $, which are the natural transverse tensors associated to the wave. Recalling the scalar field representation of the Lorentz generators $	\Sigma_{\Phi}^{\alpha \beta} =  2 i x^{[\alpha}\partial^{\beta]}$ we conclude that the two symmetries $\chi_{i}$ leaving unchanged the scalar wave $\chi_i \Phi(\phi) =0$ are
	\[
	\chi_{i} = 2i\mathcal{F}_i^{\alpha\beta}x_\alpha\partial_\beta .
	\] 
	The five Killing vector fields $(\partial_+,\partial_i, \chi_j)$ have been reported by Bondi, Pirani and Robinson in one of the first treatments of exact gravitational waves \cite{Bondi:1958aj}, along with the coordinate isometries they generate. Above, we considered a scalar wave but the same symmetries preserve electromagnetic waves as well, i.e., the vector field $A^\alpha(\phi)$. To be more precise, the Lorentz transformations $\chi_i$ in this case perform a gauge transformation on $A^\alpha(\phi)$ \cite{Kupersztych:1978, Brown:1983bc} and therefore $F_{\alpha \beta}$ is precisely invariant, we will discuss this in more detail in the following. It has been observed that these generators define a Heisenberg algebra (see e.g. \cite{Adamo:2017nia}, Eq. 2.3), which is substantially the same as the one satisfied by space and momentum operators in quantum mechanics but with $\partial_+$ replacing the identity.
	One way to extend plane waves to curved spacetime is to require the latter to exhibit at least five Killing vectors. This is what has been accomplished in Ref. \cite{Bondi:1958aj}, where it is observed that such a metric can always be locally put in a Rosen-type form $ds^2 = 2dx^+d\phi +  \gamma_{ij}(\phi)dx^idx^j$ \cite{Bondi:1957,Einstein:1937qu}. We will see that this chart exhibits coordinate singularities, however other charts describing a gravitational plane wave and covering the whole spacetime are available.
	For the sake of clarity, it is useful to discuss here in more detail the familiar case of an electromagnetic plane wave in flat spacetime $A^\alpha(\phi)$. As previously observed, this field is gauge transformed by the Lorentz-like generators $\mathcal{F}_i^{\alpha\beta}$. This being said, there is a particular combination of these generators which is the key to solve the dynamics in this background, namely $\omega^{\alpha\beta}(\phi,A) = eA^i(\phi)\mathcal{F}_i^{\alpha\beta}$ or equivalently $\omega^{\alpha\beta}(\phi,A) = e\int^\phi d \tilde{\phi} F^{\alpha\beta}(\tilde{\phi})$. As already mentioned, for generic functions $A^i(\phi)$ these define the local little group $E_2(n)$ associated to the vector $n^{\alpha}$ \cite{Brown:1983bc, Kupersztych:1978,Passarino:1986gs}. In addition, when these functions are the plane wave components it turns out that these Lorentz-like generators completely solve the motion. Namely, one can easily show that the momentum of a charged particle in a plane wave is described by \cite{Kupersztych:1978,Seipt:2017ckc}
	\[
	\pi_p^\alpha(\phi)  = \Lambda^\alpha_{ p, \beta}(\phi,A) p^\beta
	\quad ,  \qquad
	\Lambda_p(\phi,A) = \exp\left(e\int^\phi  \frac{d \tilde{\phi}}{p^-} F(\tilde{\phi})\right) \equiv e^{\frac{\omega(\phi,A)}{p^-}},
	\]
	where the transformation $\Lambda^\alpha_{p,  \beta}$ has the following properties \cite{Passarino:1986gs,Brown:1983bc,Kupersztych:1978}:
	\[
	\Lambda^\alpha_{p, \beta}\Lambda^{\gamma \beta}_{p} = \eta^{\alpha \gamma}
	\qquad ,\qquad
	\Lambda^\alpha_{p, \beta}n^\beta = n^\alpha
	\quad \Rightarrow \quad 
	\partial^\beta\Lambda^\alpha_{p, \beta} =0
	\qquad , \qquad 
	\Lambda^\alpha_{p, \beta}A^\beta = A^\alpha + \partial^\alpha f,
	\]
 with $ f(\phi)=\frac{e}{p^-}\int ^\phi d\tilde{\phi}A^{\alpha}(\tilde{\phi})A_{\alpha}(\tilde{\phi})$. The physical meaning of the above equations is that $\Lambda^\alpha_{p,  \beta}$ is a Lorentz-like transformation, it describes the Wigner little group of the null vector $n^\alpha$ and finally it defines a $U(1)$ gauge transformation. 
 The momentum evolution tells us something interesting. The motion of a particle in a plane wave can be described by local Lorentz-like transformations of the initial momentum on the constant-$\phi$ hypersurfaces. These transformations depend on the plane wave components. They do not alter the background because they act as gauge transformations and they do not change the phase direction due to the fact that they belong to its little group. These features will turn out to be very general and surprisingly useful also in the gravitational generalization. In fact, we will show that they are manifestly present in the Rosen metric. 
 The underlying connection between gauge and spacetime symmetries basically turns the problem of finding quantum field states in a plane wave background into a simple procedure. This construction will be discussed in detail later.

\section{Two complementary charts: Brinkmann and Rosen coordinates}
    \label{sec. Brinkmann and Rosen}
	There are two charts particularly useful to describe a plane wave spacetime: the Brinkmann \cite{Brinkmann:1925fr} and the Rosen \cite{Einstein:1937qu} coordinates. These metrics have the form 
	$\mathcal{G}_{\mu\nu} = \eta_{\mu\nu}
	+ H_{ij}(\phi)X^iX^j n_\mu n_\nu $ and $g_{\mu\nu} = 2 n_{(\mu} \tilde{n}_{\nu)} + \gamma_{ij}(\phi)\delta^i_\mu  \delta ^j_\nu$, respectively. While the Rosen metric shows manifestly three of the five symmetries of plane waves, it has the drawback of being not global: in general, at least two Rosen charts are needed in order to cover the whole spacetime. On the other hand, the Brinkmann chart has only one manifest symmetry but it is global and the Einstein equations have a trivial form when described in these coordinates.
    In the following, we will describe the main properties and the geodesic motion in both these charts, underlying the natural interplay between them. It is convenient to introduce from the beginning the key element connecting these metrics, namely the Rosen vierbein $e_{\alpha \mu }$ defined by 
    \[
   e_{\alpha \mu } e_{\beta \nu}\eta^{\alpha \beta} = g_{\mu\nu}.
    \]
    The matrix $e_{\beta \nu}$ will appear in the discussions of both the Rosen and the Brinkmann charts, the link between the two metrics being encoded in the evolution of its transverse part $\ddot{e}_{ij} = H_{ik}e^k_{\-\ j}$. This equation will be properly derived in the following. Note that the Rosen vierbein's first index will be assumed to be raised and lowered by the Minkowski metric while the second one by the Rosen metric. Thus, $e_{\beta \nu}$ is not a tensor in Brinkmann coordinates but just a matrix. Moreover, we will assume without losing generality the symmetry condition $\dot{e}_\alpha^{\-\  \mu }  e_{ \beta \mu} = 
	\dot{e}_\beta^{\-\  \mu }  e_{ \alpha \mu}$ \cite{Blau:2002mw}.
	
	\subsection{The Brinkmann chart}
	A plane wave is described throughout all the spacetime by the Brinkmann metric $\mathcal{G}_{\mu\nu}(X)$ \cite{Brinkmann:1925fr}
	\[
	\mathcal{G}_{\mu\nu} = \eta_{\mu\nu}
	+ H_{ij}(\phi)X^iX^j n_\mu n_\nu,  
	\]
	where $\phi = n \cdot X = T-Z $ is the same light-cone coordinate introduced in the previous sections. The independence of $\mathcal{G}_{\mu\nu}$ on $X^+ = (T+Z)/2$ makes manifest the symmetry associated to the Killing vector field $\partial_+$. It is worth noting that only one of the five symmetries of plane waves is explicit in this metric, for this reason the equations of motion in this chart are not as trivial as in the Rosen one, where three symmetries out of five are manifest. The fact that this metric is in the Kerr-Schild form assures three easily verified facts \cite{Kerr2009}: the inverse metric is simply $\mathcal{G}^{\mu\nu} = \eta^{\mu\nu}
	- H_{ij}(\phi)X^iX^j n^\mu n^\nu $, the vector $n^\mu$ is null with respects to both $\mathcal{G}_{\mu\nu}$ and $\eta_{\mu\nu}$, and, finally, the metric determinant is constant and equal to the Minkowskian one $\det \mathcal{G}_{\mu\nu} = \det\eta_{\mu\nu} = -1$. The physical information is encoded in the matrix $H_{ij}$, which is connected to the Ricci tensor by the simple relation \cite{Stephani:2003tm}
	\[
	R_{\mu\nu} = H_i^{\-\ i} n_\mu n_\nu.
	\]
	The Einstein's equations in this spacetime read
	\[
    H_i^{\-\ i} n_\mu n_\nu =- 8 \pi G T_{\mu\nu} 
    \] 
    and therefore the trace of $H_{ij}$ represents the energy density of the source which induces the gravitational perturbation. This is clearly zero for vacuum solutions, the Brinkmann profile $H_{ij}$ in this case satisfies the same traceless condition as the weak-field perturbation $h_{ij}$ in the physical transverse traceless (TT)-gauge. The pure gravitational wave is thus represented by a $2 \times 2$ traceless matrix $H_{ij}$ and we can identify the two polarizations usually introduced under the weak-field approximation
	\cite{Blau:2002mw,Harte:2012jg}
	\[
	H_{ij} = 
	\begin{pmatrix}
		H_+ & H_\times \\
		H_\times  & - H_+ 
	\end{pmatrix}.
	\]  
	Going back to the general case, where $T_{\mu\nu}\neq 0$, the weak energy condition $T_{00} \geq 0$ implies $H_i^{\-\ i} \leq 0$ \cite{Garriga:1990dp}. 
    Exploiting the simple form of the Riemann tensor it is possible to verify that its traceless part, the Weyl tensor, depends on the traceless part of the matrix $H_{ij}$.
    It thus follows that the spacetime is conformally flat if and only if the Brinkmann profile is a pure trace $H_{ij} \propto \delta_{ij}$. This is the case for an electromagnetic wave perturbating the spacetime \cite{Penrose:1965rx,Blau:2002mw,Ehlers:1962}, while pure gravitational waves are represented by the traceless components and are Ricci-flat so their Weyl tensor is equal to the Riemann one. This reflects the nature of the tidal forces acting on geodesics: in the electromagnetic case these are acting on the volume of a body while in the gravitational case they deform its shape along the two polarizations $+,\times$.
    Studying the geodesics we will see that the actual connection between the Brinkmann profile $H_{ij}$ and the usual weak-field perturbation is given by $H_{ij} = \frac{\kappa}{2} \ddot{h}_{ij} + \Ogrande(h^2)$. Despite the similarity between $H_{ij}$ and $h_{ij}$, the Brinkmann chart is not the best candidate to generalize the TT-gauge because of its dependence on the transverse coordinates, indeed the Rosen chart will be more suitable for this purpose.
	While the Einstein's equations are algebraic, the equations of motion in this metric are not completely trivial, as anticipated. The trajectories can be extracted from the Lagrangian $\mathcal{L} = m^{-1} \Pi^\mu\Pi^\nu \mathcal{G}_{\mu\nu}$, the fact that $n^\mu\partial_\mu$ is a Killing vector implies that $\Pi^-$ is conserved such that $\Pi^- = p^-$, where $p$ is the initial momentum. The transverse components are easily found to follow the harmonic equation
	\[
	\ddot{X}^i = H^i_{\-\ j } X^j.
	\]
	The last component $\Pi^+$ can be found algebraically from the on-shell condition $\Pi^\mu\Pi^\nu \mathcal{G}_{\mu\nu} = m^2$. A study of the geodesic equation shows that there are groups of null trajectories in plane wave spacetimes that converge on two nonintersecting null geodesics. This was first pointed out by Penrose \cite{Penrose:1965rx}, who deduced from this feature the impossibility to embed the spacetime into a globally hyperbolic one. A key point to study the geodesics in this chart is to find its Killing vectors. Studying the Killing equation $\mathcal{K}_{(\mu ; \nu )} = 0$, it is possible to identify two important symmetries. By introducing the matrix $e_{ij}$ satisfying $\ddot{e}_{ij} = H_{ik}e^k_{\-\ j}$ and $\dot{e}_{[i}^{\-\  j }  e_{ k] j} = 0$, the Killing vectors take the form \cite{Adamo:2017nia,Blau:2002js}
	\[
	\mathcal{K}^{\mu }_j  = \gamma_{jk}\partial^\mu e_{i}{}^{k}X^i,
	\]
	where $\gamma_{ij} = e_{li}e^l{}_{j}$ is a Rosen metric. 
    The introduction of the matrix $e_{ij}$, which as anticipated at the beginning of this section is just the transverse part of the Rosen natural vierbein, seems unjustified at this point. For now this matrix can just be considered a useful placeholder for the transverse coordinate evolution and its role will become clearer when studying the Rosen metric. From these two Killing vectors follow two conservation laws: let $\dot{X}^\mu$ be a geodesics, then the identities $\mathcal{K}_j^\mu \dot{X}_\mu $ are conserved. It is immediate to see that these are equivalent to the quantities $\partial_\phi (e_{i}{}^{j}X^i ) = c_k \gamma^{kj}$ with $c_k$ being constant. Let us now suppose that the gravitational wave belongs to the sandwich-type, such that $H_{ij}(\phi) \neq 0$ only in the interval $\phi_0 < \phi < \phi_1 $ for arbitrary fixed $\phi_{0,1}$. If we assume boundary conditions in the past flat region $\phi < \phi_0$, then we can choose $e_{ij} = \eta_{ij}$ initially and write these conservation laws as $p^-\partial_\phi( e_{i}{}^{j}X^i) = p_k \gamma^{kj} $, where $p_k$ is the particle initial momentum before the interaction with the wave. Expanding the derivative and introducing the symmetric tensor $\sigma_{ij} = \dot{e}_{i k }e_j^{\-\  k}$ we get the following expression for the transverse momentum 
	\[
 \label{P_p^i}
	\Pi^i_p = e^{ij}p_j + p^-\sigma^i_{\-\ j} X^j \equiv p^i + \Delta_p^i,
	\]
    where for future convenience we have introduced the symbol $\Delta^i_p = \Pi_p^i - p^i = \Delta e^{ij}p_j + p^-\sigma^i_{\-\ j} X^j$  to represent the correction to the transverse constant momentum one would have in flat spacetime. 
    It is worth noting that the knowledge of $\Pi^i$ still presumes the ability to solve the second-order differential equation $\ddot{e}_{ij} = H_{ik}e^k_{\-\ j}$ and the number of wave profiles $H_{ij}$ allowing to find analytical solutions is extremely limited. However, this expression can be useful to study the geodesic congruences behavior. Exploiting the on-shell condition we can write the full momentum as \cite{Adamo:2022qci, Adamo:2022rmp}
	\[
	\Pi^\mu_p =
    \eta^{\mu \alpha}p_\alpha
    + \Delta^\mu_p
    - \frac{1}{2p^-}
    \bigg[
    2 p_i \Delta^i_p + \Delta_{p,i} \Delta_p^i + (p^-)^2 H
    \bigg] n^\mu, 
	\]
	where $\Delta^\mu_p = \delta_i^\mu \Delta^i_p$ and $H = H_{ij}X^iX^j$. While by definition of plane waves $n_{\mu;\nu} = 0$, the geodesic congruence generated by $\partial_-$ has the non trivial deformation tensor $\sigma_{ij}$ \cite{Shore:2017dqx,Adamo:2017nia}. Observing that 
    \[
    \dot{\sigma}_{ij}=  H_{ij} - \sigma_{ik}\sigma_{ j}^{\-\ k}
    \]
    and exploiting the weak energy condition we can deduce by Schwarz inequality that $\dot{\sigma}_{i}^{\-\ i} + (\sigma_{i}^{\-\ i})^2 \leq 0$, which is equivalent to  $\partial^2_- \exp{\int d\phi \sigma_{i}^{\-\ i}} \leq 0$  \cite{Penrose:1965rx}. Now, if we choose the initial conditions at $\phi_0$ in an asymptotically flat region we have $\sigma_{ij}(\phi_0) = 0$, and therefore $\partial_- \exp{\int d\phi \sigma_{i}^{\-\ i}} \vert_{\phi_0} = 0$. It follows then that at some finite $\phi^*$ we will find $\exp{\int d\phi \sigma_{i}^{\-\ i}} \vert_{\phi^*} = 0$, and there some components of $\sigma_{ij}$	have to be infinite. That point is at the same time a null geodesic focusing point and a singularity of Rosen coordinates, in fact the Rosen metric $g_{\mu\nu}$ inherits the singularity from its vierbein $e_{ij}$.
    Another important property of plane wave spacetimes concerns the imprint the wave leaves on particles after its passage. These features are known as memory effects \cite{Braginsky:1985vlg, Harte:2015ila, Grishchuk_1977, Zhang:2017rno}. Let us consider a pair of particles in a sandwich-wave, which we know being Minkowskian in the in-region $\phi< \phi_0$ and out-region $\phi > \phi_1$, both initially at rest $p^\nu = (m,0,0,0)$ for simplicity. If their initial perpendicular displacement is $\Delta X_{0}^i$ then, exploiting the fact that $e_{i}^{\-\ j} \Delta X^i = \Delta X_{0}^j$ due to the symmetries discussed above Eq. (\ref{P_p^i}), we get for the momentum difference in the out-region the following expression \cite{Zhang:2017rno}
    \[
	\Delta \Pi_\nu \big \vert_{\text{out-region}} = 
	  m\dot{e}_{\nu i}(\phi_1) \Delta X_{0,\perp}^i
	- \frac{m}{2} \left( \Delta X_{0,\perp}^i \dot{e}_{j i}(\phi_1) \dot{e}^j_{\-\ k}(\phi_1) \Delta X_{0,\perp}^k \right)  n_\nu. 
	\]
    This is an example of velocity memory effect: the vierbein is trivial in the in-region $e_{ij}(\phi_0) = \eta_{ij}$ but it is constrained by the differential equation $\ddot{e}_{\nu i}(\phi_1) = 0$ in the out-region, such that its first derivative will generally be different from zero. This means that two particles initially at rest but displaced acquire a relative momentum after the passage of the wave. As we will see this property is also connected to the classical and quantum scattering of a particle by the gravitational wave itself. 
    
	\subsection{The Rosen chart}
	The other chart commonly used to describe plane wave spacetimes is the Rosen one $g_{\mu\nu}(\phi)$ \cite{Einstein:1937qu}
	\[
	g_{\mu\nu}(\phi) = 2 n_{(\mu} \tilde{n}_{\nu)} + \gamma_{ij}(\phi)\delta^i_\mu  \delta ^j_\nu,
	\]
	where $\phi$ is the same as in Brinkmann coordinates $\phi = n\cdot x$ and $\gamma_{ij}$ is a $2 \times 2$ matrix. In this chart three of the five Killing vector fields are clearly manifest: $\partial_+, \partial_i$. This will be reflected in the simplicity of the equations of motion, as we will see in the following. An important feature of this chart is that, unlike the Brinkmann one, it does not cover the whole spacetime. Indeed, it exhibits spurious coordinates singularities, as anticipated in the previous paragraph. This will become clear studying the geodesics. For this reason it is generally necessary to work in Brinkmann coordinates whenever global problems such as scattering processes are studied. Nonetheless this chart is very important in order to exploit the symmetries of plane waves and it will be essential in solving field equations. Moreover the Rosen chart can be chosen to generalize the concept of TT-gauge usually introduced in the weak-field approximation. In fact, the Rosen profile $\gamma_{ij}$ is transverse and depends on $\phi$ only, as the perturbation $h_{ij}$. The traceless property, which the Brinkmann profile shares with $h_{ij}$, is not satisfied by $\gamma_{ij}$. However, a perturbative expansion in vacuum shows that all the odd orders of $\gamma_{ij}$ are traceless while all the even ones are pure traces \cite{Harte:2015ila}. Thus, in the linear limit $\gamma_{ij} = \eta_{ij} + \kappa h_{ij}$, with the perturbation in the TT-gauge. 
    The link between the Rosen and the Brinkmann charts is easily found observing that the Brinkmann transverse coordinates, being flat, are just the Rosen vierbein projection of the Rosen ones $X^i = e^i_{\-\ j}x^j$ , with $e_{\alpha \mu } e^\alpha_{\-\ \nu} = g_{\mu\nu}$. In order to connect the profiles $\gamma_{ij}$ and $H_{ij}$ one can for example compare the Rosen and Brinkmann Riemann tensors in the Rosen chart. With the usual symmetries understood, one finds (see App. \ref{App. metrics})
	\[
	\frac{1}{4}\dot{\gamma}^{k l}\dot{\gamma}_{l m } +
	\frac{1}{2}\gamma^{k l} \ddot{\gamma}_{l m }
	=
	e^{ i k } e^{j}_{\-\ m}H_{ij}. 
	\]
	We can rewrite this relation in terms of the vierbein only as $\ddot{e}_{ij} = H_{ik}e^k_{\-\ j}$, recalling that we assumed without losing generality that the symmetry condition $\dot{e}_\alpha^{\-\  \mu }  e_{ \beta \mu} = 
	\dot{e}_\beta^{\-\  \mu }  e_{ \alpha \mu}$ is fulfilled \cite{Blau:2002mw}. This finally exhibits the connection between the Rosen and the Brinkmann charts in a clear way. For completeness we report here the full coordinate transformation that connects the two systems 
    \[
    x^- = X^- \quad , \quad 
    x^i = e_j^{\-\ i}X^j \quad , \quad 
    x^+ = X^+ + \frac{1}{2} \sigma_{ij}X^iX^j \quad , \quad
    \ddot{e}_{ij} = H_{ik}e^k_{\-\ j}.
    \]
    We recall from the previous discussion that at some focusing point $\phi^*$ the Rosen metric has a singulairty and therefore this coordinate transformation has to be singular as well in order to compensate for it and produce a globally defined Brinkmann chart.
    
	It is now time to exploit the full power of manifest symmetries in these coordinates to find the geodesics. From the three Killing vector fields $\partial_+, \partial_i$ we deduce the corresponding conserved momenta $\pi_p^- = p^-, \-\  \pi_{p,i} = p_i$. The last component $\pi^+_p$ can be derived from the on-shell condition $\pi^\mu_p \pi^\nu_p g_{\mu\nu}= m^2$. The complete momentum with initial conditions $p_\mu$ can be conveniently written as 
	\[
	\pi_{p,\mu} = p_\mu -\frac{1}{2p^-}\left(
	g^{\rho \nu}p_\rho p_\nu -m^2
	\right) n_\mu
   =
    p_\mu -\frac{p_ip_j}{2p^-}\int_{-\infty}^\phi d \tilde{\phi}\dot{\gamma}^{ij}(\tilde{\phi})       n_\mu 
    .
	\]
	It is interesting to observe that the vierbein projection of geodesics in Rosen coordinates is in a one-to-one correspondence with the motion of a charged particle in an electromagnetic plane wave. Recalling the definition $e_\alpha ^{\-\ \mu} = \delta_\alpha ^{\-\ \mu} + \Delta e _\alpha ^{\-\ \mu}$, where  $\Delta e _\alpha ^{\-\ \mu}$ is a $2 \times 2$ matrix with transverse indices, we introduce the notation 
    \[
    \Delta e_\alpha ^{\-\ \mu }p_\mu  = - \kappa P_\alpha .
    \] 
    The vierbein projected momentum $\bar{\pi}^\alpha_p = e^\alpha_{ \-\ \mu}\pi_p^\mu$ is easily found to have the form 
	\[
	\bar{\pi}^\alpha_p = 
	p^\alpha   -\kappa P^\alpha +
	\frac{\kappa}{p^-}
	\left(
	p_\beta P^\beta 
	- \frac{\kappa}{2} P_\beta P^\beta
	\right)n^\alpha,
	\]
	with $p_\alpha = \delta_\alpha^\mu p_\mu $. This is exactly the momentum of a charged particle with charge $e$ moving in an electromagnetic plane wave $A^\alpha(\phi)$ in flat spacetime, with the substitution $	\kappa P^\alpha \leftrightarrow e A^\alpha $. It is worth noting that the vector field $P^\alpha$ has only transverse degrees of freedom by definition, as the electromagnetic wave. This formal equivalence is a consequence of the symmetries shared between the Rosen metric and a plane wave in flat spacetime. In particular, one can consider the vierbein projected geodesic equation $
	p^-\dot{\bar{\pi}}_\beta + \bar{\pi}^\alpha\omega_{\alpha \beta \delta} \bar{\pi}^\delta
	= 0$, where $ \omega_{\alpha \beta \delta} $ are the spin connection coefficients $\omega_{\alpha \beta \delta} = e_\alpha^{\-\ \mu} e_\beta^{\-\ \nu}e_{\delta \nu;\mu} $. The coefficients depend on $\phi$ only, moreover the contraction $\bar{\pi}^\alpha\omega_{\alpha \beta \delta} \bar{\pi}^\delta$ is actually linear in the geodesic momentum because of the conservation of $p_i$ and $p^-$. It is easy to show that $	\bar{\pi}^\alpha\omega_{\alpha \beta \delta} = \bar{p}^\alpha\omega_{\alpha \beta \delta} =  - \kappa \mathcal{P}_{\beta\delta}$, where we introduced the analog of the plane wave Maxwell tensor $\mathcal{P}_{\beta\delta} = 2\bar{\partial}_{[\beta} P_{\delta]}$. Exploiting this result we obtain a Lorentz equation for the vierbein projected geodesics
	\[
	\dot{\bar{\pi}}^\alpha_p = \frac{\kappa}{p^-}\mathcal{P}^{\alpha\beta}\bar{\pi}_{p,\beta}.
	\]
	As before this is formally equivalent to the Lorentz equation of a particle with charge $\kappa$ in flat spacetime in presence of an electromagnetic plane wave $P^\alpha(\phi)$.
	It follows from the analysis developed in Sec. \ref{Plane_Waves} that we can write the momentum as a local Lorentz transformation of the initial one, now depending on the vierbein components
	\[\label{momentum rosen evo}
	\bar{\pi}_p^\alpha(\phi)  = \Lambda^\alpha_{ p, \beta}(\phi,P) p^\beta
	\quad ,  \qquad
	\Lambda_p(\phi,P) = \exp\left(\kappa \int^\phi _{- \infty}  \frac{d \tilde{\phi}}{p^-} \boldsymbol{\mathcal{P}}(\tilde{\phi})\right).
	\]
     The transformation $ \Lambda^\alpha_{ p, \beta}(\phi,P)$ is a gauge transformation for the vierbein, therefore $\tilde{e}_\alpha^{\-\ \mu} = e_\beta^{\-\ \mu}\Lambda^\beta_{ p, \alpha}$ defines a local reference system in which the particle has a constant momentum $p^\alpha$ throughout all the trajectory.
	This map is not only a formal interesting feature, it is very useful for translating some known results in electromagnetism and QED to systems in plane wave spacetimes and offers a very direct way to compare the linear results in these different contexts. 
    One example is the motion of a charged particle moving in an electromagnetic plane wave keeping into account the spacetime curvature produced by the latter. The vierbein projected motion in this case is described by $p^- \dot{\bar{\pi}}^\alpha = \left(\kappa\mathcal{P}^{\alpha\beta}
	+ e \bar{F}^{\alpha\beta} \right)\bar{\pi}_\beta$, therefore the momentum is formally identical to the one of a free-falling particle with the substitution $	\kappa P^\alpha \rightarrow\kappa P^\alpha +  e \bar{A}^\alpha$. While the formal analogy is as simple as clear, in the physical interpretation one has to be more careful. In fact, the connection between $P_\alpha$ and the physical complete gravitational wave profile $H_{ij}$ is $\ddot{P}_i = H_{i}^{\-\ j}( p_j + P_j)$. Nonetheless, in the linear limit it is $P_\alpha = \frac{1}{2}p^\beta h_{\beta \alpha}$ and this is the most natural representative of the gravitational perturbation in the weak-field limit.

	\section{Quantum fields in a plane wave } \label{sec. QF}
	\subsection{In flat spacetime}
	In Sec. \ref{Plane_Waves} we noticed that in an electromagnetic background the Lorentz-like transformation generated by $\omega^{\alpha\beta}(\phi,A) = e\int^\phi d \tilde{\phi} F^{\alpha\beta}(\tilde{\phi})$ plays a very important role. Indeed, $\Lambda_p = \exp \left(  \omega/p^- \right)$ completely solves the dynamics acting as an evolution operator on the initial four-momentum:
	$
	\pi_p^\alpha  = \Lambda^\alpha_{ p, \beta} p^\beta
	$.
	The importance of this transformation is not limited to particle dynamics but it is essential in solving fields equations. 
    Let us introduce the Hamilton-Jacobi action 
    \[
    S_p(x) = -p \cdot x
    - \frac{e}{p^-}
    \int^\phi d \tilde{\phi}
    \left(
    p\cdot A - \frac{e}{2} A\cdot A \right)
    \]
    such that 
    \[e^{-iS_p}iD_{\alpha}e^{iS_p}=e^{-iS_p}(i\partial_{\alpha}-eA_{\alpha})e^{iS_p} = \pi_{p,\alpha}.
    \]
    From the momentum evolution Eq. (\ref{momentum rosen evo}) we find that $e^{-iS_p}iD_{\alpha}e^{iS_p} = \Lambda^\alpha_{ p, \beta} p^\beta$. This means that the quantum operator $\mathcal{U} = \exp i \left(S_p + p\cdot x\right)\vert_{p \rightarrow i \partial}$ constructed from the difference between the dressed and the free action is a unitary transformation that changes the covariant derivative into the locally Lorentz transformed free derivative \cite{Brown:1983bc, Kupersztych:1978}. If we define the operator $\Lambda = \Lambda_p\vert_{p \rightarrow i \partial}$ we find the useful relation
	\[
	\mathcal{U}^{-1} D^\alpha \mathcal{U}  = \Lambda^\alpha_{ \-\ \beta} \partial^\beta.
	\]
	This can be seen as an evolution operator for the free field $\varphi_p = e^{- i p\cdot x}$ such that $\varPhi_p = \mathcal{U} \varphi_p$ solves the wave equation $\left( D^\alpha D_\alpha + m^2 \right) \varPhi_p= 0$. It is immediate to check this exploiting the identity $\mathcal{U}^{-1} D^2 \mathcal{U}  =  \partial^2$, which in turn follows from the properties $\Lambda^\alpha_{\-\ \beta}\Lambda^{\gamma \beta} = \eta^{\alpha \gamma}
    $ and $
	\Lambda^\alpha_{\-\ \beta}n^\beta = n^\alpha$, recalling that $\Lambda^{\alpha}{}_{\beta}$ depends only on $\phi$. 
    We now proceed considering Dirac spinors and massive charged vector fields \cite{ Brown:1983bc, Nikishov:2001fd}, as we will see there is a very natural way to exploit $\Lambda$ in order to find these quantum states.
    Let us introduce the generalized quantum operator
	\[
	\mathbf{\Lambda}(\phi,A) = e^{\frac{ e }{2 \partial_+}\int^\phi d \tilde{\phi} F^{\alpha\beta}(\tilde{\phi}) \mathbf{\Sigma}_{\alpha\beta} },
	\] 
	where now $\mathbf{\Sigma}_{\alpha\beta}$ are the generators of the Lorentz algebra in a general representation. Above, the four-vector representation was considered, where $(\mathbf{\Sigma}_{\alpha\beta})^{\gamma\delta}=-i(\delta_{\alpha}{}^{\gamma}\delta_{\beta}{}^{\delta}-\delta_{\alpha}{}^{\delta}\delta_{\beta}{}^{\gamma})$. The wave equations for fields $\boldsymbol{\varPhi}_p$ of spin $\left\{0,\frac{1}{2},1\right\}$ in an electromagnetic plane wave can be written as 
	\[(\tilde{\nabla}^\alpha \tilde{\nabla}_\alpha + m^2) \boldsymbol{\varPhi}_p = 0
	\qquad , \quad \text{with} \qquad
	\tilde{\mathbf{\nabla}}_\alpha  = 
	D_\alpha\mathbf{1} - \frac{e}{2 \partial_+} n_\alpha F^{\gamma\beta }\mathbf{\Sigma}_{\gamma\beta} 
	= \mathbf{\Lambda} D_\alpha \mathbf{\Lambda}^{-1}.  
	  \] 
	Note that in order to prove the last equality one has to exploit the identity $[F^{\alpha\beta}(\phi)\mathbf{\Sigma}_{\alpha\beta},F^{\gamma\delta}(\phi')\mathbf{\Sigma}_{\gamma\delta}]=0$ for arbitrary values of $\phi$ and $\phi'$, which derives from the fact that $\mathbf{\Sigma}_{\alpha\beta}$ are the generators of the Lorentz algebra. One can expand $\tilde{\nabla}^\alpha \tilde{\nabla}_\alpha = D_\alpha D^\alpha - e F^{\gamma\beta }\mathbf{\Sigma}_{\gamma\beta}  $ and check that this reproduces the equations studied for example in \cite{Brown:1983bc,Nikishov:2001fd} for the scalar, Dirac and vector fields
	\[
	&(D_\alpha D^\alpha + m^2 ) \varPhi_p = 0, 
    \\
    &(D_\alpha D^\alpha + i e \slashed{n}\slashed{A}  + m^2 ) \varPsi_p = 0,
    \\
    &[ (D_\alpha D^\alpha + m^2)\delta^\beta_\gamma + 2ie F^\beta_{\-\ \gamma} ]\varPhi^\gamma_p = 0 
    \quad , \quad D_\alpha \varPhi^\alpha_p = 0. 
	\]
    Making use of the definition $\tilde{\mathbf{\nabla}}_\alpha   
	= \mathbf{\Lambda} D_\alpha \mathbf{\Lambda}^{-1} $, we can write the equations $(\tilde{\nabla}^\alpha \tilde{\nabla}_\alpha + m^2) \boldsymbol{\varPhi}_p = 0$ as $(D^\alpha D_\alpha + m^2) \mathbf{\Lambda}^{-1} \boldsymbol{\varPhi}_p = 0$, such that the problem is reduced to finding the scalar field solution. This problem has already been solved with $\varPhi_p = \mathcal{U} \varphi_p$ and we thus conclude that $\mathcal{U} \mathbf{\Lambda}$ is the evolution operator for fields up to spin one
	\[
	\boldsymbol{\varPhi}_p =  \mathbf{\Lambda} \mathcal{U} \boldsymbol{\varphi}_p,
	\]
	where $\boldsymbol{\varphi}_p$ is the initial condition in absence of the background wave. This clearly corresponds to the fact that $\mathbf{\Lambda}$ satisfies the equation 
	$
	\mathbf{\Lambda}^{-1} \mathcal{U}^{-1}\tilde{\mathbf{\nabla}}^\alpha \mathcal{U} \mathbf{\Lambda}
	=
	\Lambda^\alpha_{ \-\ \beta} \partial^\beta 
	$ and from this it follows $
	\mathbf{\Lambda}^{-1} \mathcal{U}^{-1}\tilde{\mathbf{\nabla}}^2 \mathcal{U} \mathbf{\Lambda}
	=
	 \partial^2 
	$. For the sake of clarity, let us introduce the $\phi$-dependent polarization $\mathbf{S}_p(\phi)$ such that 
    \[
    \boldsymbol{\varPhi}_p =   \mathbf{S}_p e^{iS_p}.
    \]
    Defining $\mathbf{\Lambda}_p = \mathbf{\Lambda}\vert_{\partial \rightarrow -ip}$ it is easy to check that this can be explicitly written as $\mathbf{S}_p = \mathbf{\Lambda}_p \bm{s}_p$, where $\bm{s}_p = \{  1, u_p ,\varepsilon_p^\alpha \}$ are the free polarizations for scalar, Dirac, and vector fields, respectively. For example, in the well-known Volkov solution \cite{Volkov_1935} $\psi_p = U_p e^{iS_p}$ the spinorial term can be found as $U_p^a =  \bm{\Lambda}^a_{p , b}u_p^b$, such that
    \[
    \psi_p  =  e^{iS_p}\left( 1  + e\frac{\slashed{n}\slashed{A}}{2 p^-}      \right)u_p. 
    \]
    In the same way, the spin-one solution $\varPhi_p^\alpha = \mathcal{E}^\alpha_p e^{iS_p}$ is found to be \cite{Brown:1983bc}
    \[
    \varPhi_p^\alpha =e^{iS_p}
    \left[
    \varepsilon^\alpha_p   -  \frac{ \varepsilon^-_p}{p^-} e A^\alpha +
	\frac{e}{p^-}
	\left(
	\varepsilon_p^\beta A_\beta 
	- \frac{ \varepsilon^-_p}{2 p^-} e A_\beta A^\beta
	\right)n^\alpha
     \right].
    \]
    Being in the vector representation, $\mathcal{E}^\alpha_p = \Lambda^\alpha_{p , \beta} \varepsilon^\beta_p$ reduces to the particle classical momentum if $\varepsilon_p^\alpha$ is replaced by $p^\alpha$. The Lorenz condition corresponds to the natural requirement $\mathcal{E}^\alpha_p \pi_{p,\alpha} = \varepsilon_p^\alpha p_\alpha = 0$.
    \\
    As a side note, let us mention that the symbol $\tilde{\mathbf{\nabla}}_\alpha$ was not randomly chosen. In fact, it defines a sort of parallel transport rule for the field. From the properties of the operator $\tilde{\nabla}_\alpha$ it follows that $\tilde{\nabla}^\alpha \boldsymbol{\varPhi}_p = - i \Lambda_{p,\beta}^{\alpha} p^\beta \boldsymbol{\varPhi}_p $. Thus, a contraction with the momentum gives $\pi_p^\alpha\tilde{\nabla}_\alpha \boldsymbol{\varPhi}_p = - i m^2 \boldsymbol{\varPhi}_p$. If we insert the expression $\boldsymbol{\varPhi}_p =   \mathbf{S}_p e^{iS_p}$ this equation reduces to a condition on the polarization function, which is formally equivalent to a parallel transport 
    $
    \pi_p^\alpha\tilde{\nabla}'_{p,\alpha} \mathbf{S}_p = 0
    $, where $\tilde{\nabla}'_{p,\alpha} =  \partial_\alpha\mathbf{1} - i\frac{e}{2 p^-} n_\alpha F^{\gamma\beta }\mathbf{\Sigma}_{\gamma\beta}  =  \mathbf{\Lambda }_p\partial_\alpha \mathbf{\Lambda}^{-1}_p$. It is also worth noting that if we choose the vector representation then $    \pi_p^\alpha\tilde{\nabla}'_{p,\alpha} \pi_p^\beta = 0$ is actually equivalent to the Lorentz equation.
	We could rephrase the preceding discussion as follows: to construct the fields previously considered one can first operate a unitary transformation $\mathcal{U}$ that turns the gauge derivative $D_\alpha$ into $\Lambda^\alpha_{ \-\ \beta} \partial^\beta$, then transport the polarization function through the operator $\tilde{\nabla}'_{p,\alpha}$. 
    We will now see how these considerations can be translated in a plane wave spacetime.

    \subsection{In the Rosen metric }
	The Rosen chart is a particularly appropriated choice to solve fields equations. Let us start with the scalar field satisfying $(\nabla_\mu \nabla^\mu + m^2) \Phi = 0$. In vacuum flat spacetime we are used to expand fields in terms of momentum eigenstates but in a general curved spacetime this is not possible because there is no clear global notion of Fourier components. Nonetheless, the plane wave spacetime is quite peculiar because in this space the scalar wave satisfies the Huygens' principle \cite{Friedlander:1975,Ward:1987ws}. 
 By exploiting the identity $\nabla_\mu \nabla^\mu \Phi = \inv{\sqrt{g}}\partial_\mu (\sqrt{g}\partial^\mu\Phi)$ we can rewrite the wave equation as $( \dot{\log}(\sqrt{g})\partial_+ + \partial^\mu \partial_\mu + m^2) \Phi = 0$. Now, being $\partial_i, \partial_+$ symmetries of the spacetime under consideration, we will assume the solution to be an eigenstate of these operators. It is then easy to reduce the equation to 
	\[
	( \partial^\mu \partial_\mu + m^2)  g^{\frac{1}{4}}\Phi = 0
	.\]
	From this equation one immediately sees that $g^{\frac{1}{4}}\Phi_p = e^{iS_p}$ is a solution with initial momentum $p$ if $S_p$ is the action satisfying $g^{\mu\nu}\partial_\mu S_p \partial_\nu S_p= m^2$ and $\partial^\mu \partial_\mu S_p = 0$. Introducing the notation $\Omega =g^{-\frac{1}{4}} $ we can write the scalar wave as
	\[\Phi_p = \Omega e^{iS_p}.\]
	The action can be easily computed exploiting the symmetries of the metric, but it can also be borrowed from electromagnetism. If we project the derivatives onto the vierbein and we split it as in the previous section $e_\alpha ^{\-\ \mu} = \delta_\alpha ^{\-\ \mu} + \Delta e _\alpha ^{\-\ \mu}$ we obtain 
	$	[  \eta^{\alpha \beta}( \partial_\alpha + \Delta e _\alpha ^{\-\ i} \partial_i)( \partial_\beta + \Delta e _\beta ^{\-\ j} \partial_j) + m^2]  \Omega^{-1}\Phi_p = 0
	$. By observing that $\eta^{\alpha \beta} [\Delta e _\alpha ^{\-\ \mu} \partial_\mu,  \partial_\beta + \Delta e_\beta ^{\-\ \nu} \partial_\nu] = 0$
	and by recalling that $\Phi_p$ is eigenstate of $\partial_i$, this equation can be written as 
	\[	
	[  \eta^{\alpha \beta}( \partial_\alpha + i\kappa P_\alpha)( \partial_\beta + i\kappa P_\beta) + m^2]  \Omega^{-1}\Phi_p = 0.
	\]
	Now, this is formally the scalar wave equation for the field $\Omega^{-1}\Phi_p$ in an electromagnetic plane wave with the potential and charge product replaced by $\kappa P_\beta$. It then follows that the action $S_p(x)$ is just the one of a charged particle moving in an electromagnetic plane wave under the same substitution: 
	\[
	S_p(x) = -p_\mu  x^\mu  
	- \frac{\kappa}{p^-}
	\int^\phi d \tilde{\phi} 
	\left(
	p_\alpha P^\alpha - \frac{\kappa}{2}P_\alpha P^\alpha	\right). 
	\]
	To summarize, we found that the solution of the Klein-Gordon equation in the Rosen metric is $ \Phi_p = \Omega \varPhi_p$, where $\varPhi_p$ is formally the field with charge $\kappa$ satisfying the Klein-Gordon equation in flat spacetime with an electromagnetic plane wave potential $P_\beta$.
   
    The same is true for Dirac fields, such that the spinor in a plane wave spacetime is $\Psi_p =\Omega\varPsi_p  $, where $\varPsi_p$ is the Volkov state \cite{Volkov_1935} of charge $\kappa$ in a wave potential $ P_\beta$
	\[
	\Psi_p  = \Omega e^{iS_p} \left( 1  + \kappa \frac{\slashed{n}\slashed{P}}{2 p^-}      \right)u_p. 
	\]
	This can be verified by observing that the Dirac equation $(i \gamma^\mu \nabla_\mu -m) \Psi_p = 0 $ can be reduced to $[ i (\slashed{\partial}+i \kappa \slashed{P})  - m ]\Omega^{-1} \Psi_p  $ and this is once again formally the Dirac equation in flat spacetime in presence of a plane wave field $\kappa P^\alpha$. In this case, the spin connection term $\gamma^\mu \Gamma_\mu$ is exactly compensated by the $\Omega$ prefactor.
 
    The spin-one massive case is slightly more involved. Let us introduce the notation $\bar{A}^\alpha = \Omega \bar{\varPhi}^\alpha$, where $\bar{A}^\alpha = e^\alpha_{\-\ \mu} A^\mu$ is the vierbein-projected vector field. Keeping into account the condition $\nabla_\mu A^\mu = 0 $ it is possible to reduce the field equation $( \nabla_\mu \nabla^\mu + m^2) A^\nu 
    + R^\nu_{\-\ \mu}A^\mu
    = 0  $ to the form 
    \[
    \label{Vector_Field_Rosen_1}
    (\bar{\partial}^\beta \bar{\partial}_\beta 
	+ m^2
	)\bar{\varPhi}^\alpha  + 2 i \kappa \hat{\mathcal{P}}^{\alpha}_{\-\ \gamma} \bar{\varPhi}^\gamma  
	+ 
	\frac{1}{2} \ddot{\log}g \bar{\varPhi}^- n^\alpha = 0,
    \]
    where we introduced the quantum analog of the Maxwell-like tensor $\mathcal{P}_{\alpha \beta}$ substituting $p_i \rightarrow i \partial_i$, namely $ \kappa \hat{\mathcal{P}}^{\alpha \beta} = -\mathcal{F}_j^{\alpha \beta} \dot{e}^{jk}i\partial_k$. 
    Here, we see that the last term in the l.h.s., which comes from the contraction of two spin connections, is a nonlinear term absent in flat spacetime. However, contracting with $n^\alpha$ one finds that $(\bar{\partial}^\beta \bar{\partial}_\beta 
	+ m^2
	)\bar{\varPhi}^- = 0 $
    and therefore assuming initial momentum $p$ one finds $\bar{\varPhi}^-_p = \varepsilon^-_p e^{iS_p}$, with $\varepsilon^-_p$ constant. Exploiting the property $\bar{\partial}_\beta \bar{\varPhi}^-_p = -i \bar{\pi}_{p,\beta} \bar{\varPhi}^-_p  $, one can check that the shift 
	$\bar{\varPhi}'^\alpha_p = \bar{\varPhi}^\alpha_p + \frac{i}{4 p^-}\dot{\log}g\bar{\varPhi}^-_p n^\alpha $ restores the form of a flat spacetime field equation in an electromagnetic wave background 
	\[
	(\bar{\partial}^\beta \bar{\partial}_\beta 
	+ m^2
	)\bar{\varPhi}'^\alpha_p  + 2 i \kappa  \mathcal{P}^{\alpha}_{\-\ \gamma} \bar{\varPhi}'^\gamma _p 
	= 0.
	\] 
    Indeed, this field satisfies the condition $\bar{\partial}_\alpha\bar{\varPhi}_p'^\alpha = ( \partial_\alpha + i \kappa P_\alpha)\bar{\varPhi}'^\alpha_p = 0 $, which is the analog of $D_\alpha \varPhi^\alpha_p = 0$ found in flat spacetime in the previous section.

    Finally, we can report the solution for the massive vector field in the form $\bar{A}_p^\alpha = \Omega e^{iS_p}\bar{\mathcal{E}}^\alpha_p $, namely
    \[
    \bar{A}^\alpha_p = \Omega e^{iS_p}
    \left[
    \varepsilon^\alpha_p   -  \frac{ \varepsilon^-_p}{p^-} \kappa P^\alpha +
	\frac{\kappa}{p^-}
	\left(
	\varepsilon_p^\beta P_\beta 
	- \frac{ \varepsilon^-_p}{2 p^-} \kappa P_\beta P^\beta
	\right)n^\alpha
    + i\dot{\log}{\Omega}\frac{ \varepsilon^-_p}{p^-}  n^\alpha
     \right].
    \]
    Here one can explicitly observe once again the formal analogy with the case of an electromagnetic background in flat spacetime. Once the shift previously described is operated the polarization $\bar{\mathcal{E}}^\alpha_p$ can be found substituting $e A^\alpha \rightarrow \kappa P^\alpha$ in the flat spacetime solution. If the massless case is considered one can exploit the gauge freedom to fix $\bar{\partial}_\alpha \bar{\varPhi}^\alpha_k = 0 = \bar{\Phi}^-_k$, which implies $\nabla_\mu A_k^\mu = 0 $ as well. In this case the additional term proportional to $\ddot{\log}g$ in Eq. (\ref{Vector_Field_Rosen_1}) vanishes and the solution is once again in direct correspondence with the one in flat spacetime without any shift needed. We will deal with this choice studying the field in Brinkmann coordinates. 
    The massive spin-two field is not of our interest in the present work and we will confine ourselves to the massless case, which is easily treated in Brinkmann coordinates once the proper gauge is chosen \cite{Adamo:2017nia}. 
    
	We can now retrace the discussion carried out in flat spacetime, i.e., that the operator constructed from the difference between the free and the curved spacetime actions $\mathcal{U} = \exp i \left(S_p + p_\mu x^\mu \right)\vert_{p \rightarrow i \partial}$ is a unitary transformation that changes the vierbein projected derivative into the Lorentz transformed curved one $\mathcal{U}^{-1}\bar{\partial}_\alpha \mathcal{U} = \Lambda_\alpha^{\-\ \beta}\partial_\beta $, with $\partial_\alpha = \delta_\alpha^\mu \partial_\mu $.
    We can introduce the gravitational analog of the Lorentz-like transformation exploited in the electromagnetic wave background treatment, which can be obtained through the simple substitution emerged in the study of the dynamics $e\dot{A}^j \rightarrow -i\kappa\Delta \dot{e}^{jk}\partial_k$
    \[
    \mathbf{\Lambda} = 
    e^{\frac{ \kappa }{2 \partial_+}\int^\phi _{ -\infty} d \tilde{\phi} \hat{\mathcal{P}}^{\alpha\beta}(\tilde{\phi}) \mathbf{\Sigma}_{\alpha\beta} }.
    \]     
    The wave equation for the fields $\{\Phi, \Psi, \bar{A}'^\alpha \} \equiv  \bar{\boldsymbol{\Phi}}_p $ can be written as
	\[
	(\tilde{\nabla}^\alpha \tilde{\nabla}_\alpha + m^2) \Omega^{-1}\bar{\boldsymbol{\Phi}}_p = 0
	\qquad , \quad \text{with} \qquad
	\tilde{\mathbf{\nabla}}_\alpha  = 
	\bar{\partial}_\alpha\mathbf{1} - \frac{\kappa}{2 \partial_+} n_\alpha \hat{\mathcal{P}}^{\gamma\beta }\mathbf{\Sigma}_{\gamma\beta} = 
    \mathbf{\Lambda} \bar{\partial}_\alpha \mathbf{\Lambda}^{-1} , 
	\] 
    where we recall that the spin-one field involves the shift $\bar{A}'^\alpha = \bar{A}^\alpha - \frac{i}{ p^-}\dot{\log}\Omega\bar{A}^- n^\alpha $.
    \\
	Let us point out that $\tilde{\nabla}_\alpha$ is not the vierbein projected covariant derivative. This operator is defined by the representation of the considered field and not by the object it is acting on. This is where its relevance comes from as it allows one to write the wave equation as the actual square of an operator.
	However, when taken along a geodesic this operator is actually equivalent to the vierbein-projected covariant derivative for the representations here discussed, namely $\bar{\pi}^\alpha_p \tilde{\nabla}_{p,\alpha} = \bar{\pi}^\alpha_p \bar{\nabla}_{\alpha}$, with $\tilde{\mathbf{\nabla}}_{p,\alpha}  = 
	\bar{\partial}_\alpha\mathbf{1}  - i \frac{\kappa}{2 p^-} n_\alpha \mathcal{P}^{\gamma\beta }\mathbf{\Sigma}_{\gamma\beta}  $. This is quickly verified exploiting the spin coefficients property $	\bar{\pi}^\alpha\omega_{\alpha \beta \delta}= - \kappa \mathcal{P}_{\beta\delta}$. 
	To complete the correspondence we have the properties
	\[
	\mathbf{\Lambda}^{-1} \mathcal{U}^{-1}\tilde{\mathbf{\nabla}}_\alpha \mathcal{U} \mathbf{\Lambda}
	=
	\Lambda_\alpha^{ \-\  \beta} \partial_\beta 
	\qquad , 
	\qquad 
	\mathbf{\Lambda}^{-1} \mathcal{U}^{-1}\tilde{\mathbf{\nabla}}^2 \mathcal{U} \mathbf{\Lambda}
	=
	\eta^{\alpha \beta}\partial_{\alpha} \partial_\beta. 
	\]
	This allows us to easily find the aforementioned fields as
	\[
	\bar{\boldsymbol{\Phi}}_p =  \Omega\mathbf{\Lambda} \mathcal{U} \boldsymbol{\varphi}_p,
	\]
    where $\boldsymbol{\varphi}_p$ is the flat initial condition. Once again we can introduce the polarization $\mathbf{S}_p(\phi)$ such that $\bar{\mathbf{\Phi}}_p = \Omega\bar{\mathbf{S}}_p(\phi)e^{iS_p}$.
    As before one can check that this can be written as $\bar{\mathbf{S}}_p = \mathbf{\Lambda}_p \bm{s}_p$ where $\bm{s}_p = \{  1, u_p ,\varepsilon_p^\alpha \}$ are the flat-spacetime free polarizations for scalar, Dirac and vector fields respectively.
    \\
	By adapting the discussion in an electromagnetic wave background we find that
	$\tilde{\mathbf{\nabla}}_\alpha \Omega^{-1} \bar{\mathbf{\Phi}}_p  = - i \Lambda_\alpha^{ \-\  \beta} p_\beta \Omega^{-1} \bar{\mathbf{\Phi}}_p $ and from this  $\bar{\pi}^\alpha\tilde{\mathbf{\nabla}}_\alpha \Omega^{-1} \bar{\mathbf{\Phi}}_p  = - im^2 \Omega^{-1} \bar{\mathbf{\Phi}}_p$. Inserting into this equation the form $\bar{\mathbf{\Phi}}_p = \Omega\bar{\mathbf{S}}_p(\phi)e^{iS_p}$ we obtain a condition for the polarization which reads $\bar{\pi}_p^\alpha \tilde{\nabla}'_{p,\alpha} \bar{\mathbf{S}}_p = 0$, where $\tilde{\mathbf{\nabla}}'_{p,\alpha}  = 
	\partial_\alpha\mathbf{1} - i \frac{\kappa}{2 p^-} n_\alpha \mathcal{P}^{\gamma\beta }\mathbf{\Sigma}_{\gamma\beta}$. Now, observing that $\bar{\pi}_p^\alpha \tilde{\nabla}'_{p,\alpha} \bar{\mathbf{S}}_p = \bar{\pi}_p^\alpha \tilde{\nabla}_{p,\alpha} \bar{\mathbf{S}}_p  =
    \bar{\pi}_p^\alpha \bar{\nabla}_{p,\alpha} \bar{\mathbf{S}}_p $, we finally find that the polarization function is defined by a proper parallel transportation $ \pi^\mu_p \nabla_\mu \mathbf{S}_p = 0$, or
	\[
	\frac{D \mathbf{S}_p}{D\phi} = 0.
	\]
	This is a covariant equation and it holds in other charts as well. In particular, it can be used as an alternative way to find the states in Brinkmann coordinates. Moreover, this equation underlines once again the semiclassical behaviour of quantum states in a plane wave background and offers an intuitive geometric picture. For example, the polarization of the shifted spin-one field evolves as a free falling gyroscope \cite{Weinberg_1972} being just parallel transported. 

	\subsection{In the Brinkmann metric}
    It is needless to say that we could just transform the fields obtained in the Rosen chart and find the states in Brinkmann coordinates. However here we want to underline the role of the operator $\mathbf{\Lambda}$ in this chart and also the form this takes in a particular gauge reproducing the spin-raising operator studied in \cite{Adamo:2017nia} for massless fields. 
    The scalar field can be clearly expressed as $\Phi_p = \Omega e^{iS_p}$ where $S_p$ is the classical action for a particle in Brinkmann coordinates \cite{Adamo:2017nia}
    \[
	S_p(X) = - p^-X^+ - X^i e_i^{\-\ j}p_j   -
	\frac{p^-}{2}\sigma_{ij}X^i X^j 
    + \frac{1}{2 p^-} \int^{\phi} d \tilde{\phi} (p_ip_j\gamma^{ij} - m^2).
	\]
	The operator $\mathbf{\Lambda}$ can be written in this chart as
	\[
	\mathbf{\Lambda} = e^{-\frac{ i }{2 \partial_+}\mathbf{\Sigma}_{\alpha \beta} \mathcal{F}_i^{\alpha \beta} \eta^{ij}(\partial_{X^j}-  e^k_{\-\ j}\partial_{X^k} )}.
	\] 
	If we choose the natural vierbein associated to this chart $E_{\alpha \mu} = \eta_{\alpha\mu} + \frac{H}{2}n_\alpha n_\mu$
	we can immediately find the spinor solutions, in fact $E_{ij} = e'_{ij}$  where $e'_{ij} = e_{ik} \pder{x^k}{X^j} $ is the Rosen transverse vierbein in Brinkmann coordinates. We can thus expand the operator to obtain 
	\[
	\Psi_p = 
	\Omega e^{iS_p}\left(
	1 + \frac{\slashed{\Delta}_p\slashed{n} }{2p^-} ,
	\right)u_p,
	\]
	where we recall that $\Delta^i_p = \Pi_p^i - p^i$. Once again the solution is similar to the Volkov one, with the substantial difference that now $\Delta^i_p$ depends on the three coordinates $\phi,X^i$ and not just on $\phi$. This will be a major difference as compared to the electromagnetic case when we will calculate $S$-matrix elements, along with the definition of in- and out-states discussed in the next section. 
    This solution can be verified to match the one found in the Rosen chart, in fact the transformation of the spinor field is generated by the Lorentz transformation connecting the Brinkmann chosen vierbein and the Rosen one expressed in Brinkmann coordinates $E^\alpha_{\-\ \mu}  = \tilde{\Lambda}^\alpha_{\-\ \beta} e^\beta_{\-\ \nu}  \pder{x^\nu}{X^\mu} = 
	\tilde{\Lambda}^\alpha_{\-\ \beta} e'^\beta_{\-\ \mu} $. Expanding $	\tilde{\Lambda}_{\alpha \beta} =\eta_{\alpha \beta } -\omega_{\alpha\beta}$ the infinitesimal Lorentz transformation is found to be $\omega_{\alpha \beta} = 2 n_{[\alpha} \sigma_{\beta] i} X^i $ and accordingly the transformed spinor follows as 
    \[
    \Psi_p = 
    e^{-\frac{i}{2}\omega_{\alpha\beta}\bm{\Sigma}^{\alpha \beta}} 
    \Omega e^{iS_p} \left( 1  + \kappa \frac{\slashed{n}\slashed{P}}{2 p^-}      \right)u_p = 
	  \Omega e^{iS_p} 
    e^{\frac{1}{2} \sigma_{\alpha i} X^i \bar{\gamma}^\alpha  \slashed{n}} 
   \left( 1  + \kappa \frac{\slashed{n}\slashed{P}}{2 p^-}      \right)u_p
    =
	\Omega e^{iS_p}\left(
	1 + \frac{ \slashed{\Delta}_p\slashed{n}}{2p^-} 
	\right)u_p.
    \]
    One can also verify that the spinorial matrix satisfies the parallel transport equation, as discussed in the context of the Rosen metric. 
    Let us now consider the photon field. If we fix the gauge such that only the transverse components of the flat polarization are left $\varepsilon^\mu= \delta^\mu_i \varepsilon^i$ and they obey the Lorentz condition $k_i\varepsilon^i = 0$, we can reduce the operator action to 
	$
	A_k^\mu =\Omega\mathbf{\Lambda}^\mu_{\-\ i}\varepsilon^i e^{iS_k} 
	\equiv 
	\Omega R^\mu e^{iS_k}
	$,
	where we introduced the spin raising operator studied in \cite{Adamo:2017nia, Mason1989ward} $R^\mu = \mathbf{\Lambda}^\mu_{\-\ i} \varepsilon^i = 
	\frac{1}{\partial_+}\left[
	\delta_i^\mu \partial_+ -
	\partial_i n^\mu 
	\right]\varepsilon^i =
	-\frac{1}{\partial_+}\varepsilon^i \mathcal{F}^{\mu \nu}_i\partial_\nu$.
    In this gauge the massless vector field is thus 
    \[
    A_k^\mu  =
	\Omega e^{iS_k} 
	\left(		
	\delta_i^\mu -
	\frac{ 1 }{ k^-}  \Delta_{k,i} n^\mu 	
	\right)\varepsilon^i.
    \]
    This is clearly in agreement with the field found in the Rosen chart once the same gauge is imposed.
    \\
    As noticed in \cite{Adamo:2017nia}, in vacuum we can choose for a massless spin-2 field the covariant TT-gauge and fix $n^\mu h_{\mu\nu} = 0$. As a result the field can be found applying two times the spin raising operator, such that  $h_k^{\mu \nu} = \Omega R^\mu R^\nu \Phi_k$. The polarization tensor is easily found to be 
	$
	\mathcal{E}_k^{\mu\nu}  = \mathcal{E}_k^\mu \mathcal{E}_k^\nu
	- \frac{i}{k^-} \sigma_{ij}\varepsilon^i \varepsilon^j n^\mu n^\nu
	$ 
	such that 
    \[
    h_k^{\mu\nu} =
			\Omega e^{iS_k}
			\left[		
			\left(
			\delta_i^\mu -
			\frac{ \Delta_{k,i}n^\mu }{ k^-}   	
			\right)
			\left(
			\delta_j^\nu -
			\frac{ \Delta_{k,j}n^\nu }{ k^-}   	
			\right)
			- \frac{i}{k^-}\sigma_{ij}n^\mu n^\nu 
			\right]
			\varepsilon^i\varepsilon^j.
    \]

	\section{Compton scattering }
 \label{sec. Compton}
	\subsection{Definition of scattering states}
	When dealing with scattering problems we have to work in Brinkmann coordinates, because the regularity of fields over the whole spacetime is essential.
    Due to the presence of a gravitational wave in- and out-states are generally different. Let us consider for simplicity the scalar case $\Phi(X) = \Omega(\phi) e^{iS(X)}$. Moreover we will assume the spacetime to be flat at $\phi = \pm \infty$. We can impose the free plane wave boundary condition in the in- or out-region and define the positive energy in- and out-states, respectively $\Phi^\uparrow, \Phi^\downarrow$ as \cite{Garriga:1990dp,Adamo:2017nia}
	\[
	\Phi_p^{\updownarrow} = \Omega^\updownarrow e^{iS_p^\updownarrow} 
	\qquad , 
	\qquad
	\lim_{\phi \rightarrow \mp \infty}\Phi_p^{\uparrow,\downarrow} = e^{-ip\cdot X}.
	\] 
	As observed in the previous sections, these fields depend on a matrix $e_{\alpha \mu}$, which is also the vierbein of a Rosen metric. This is the reason why in- and out-states are different, they depend on vierebeins which do not represent the physical wave profile $H_{ij}$ but they are connected to it through the differential equation $\ddot{e}_{i k} = H_{ij}e^j_{\-\ k}$. A vierbein does not have to be trivial $e_{\alpha \mu} = \eta_{\alpha \mu}$ in order to describe a flat region, but it can be at most linear in $\phi$ such that $\ddot{e}_{i k} = 0$. In other words a Rosen metric does not have to be Minkowskian in a flat region and if we require it to be so, for example in the in-region, it will not be Minkowskian in the out-region. The boundary conditions are thus transferred to the vierbeins, such that 
	\[
	\lim_{\phi \rightarrow \mp \infty} (e^{ \uparrow, \downarrow} )_{\-\ \mu}^\alpha = \delta^\alpha _\mu
	\qquad , \qquad
	(e^{ \uparrow, \downarrow} )_{\-\ \mu}^\alpha = (b^{\uparrow, \downarrow } )^\alpha_{\-\ \mu} \phi + 
    (c^{\uparrow, \downarrow } )^\alpha_{\-\ \mu}
	\quad \text{for} \quad \phi \rightarrow \pm \infty.
	\]
	There are only two constrains on these coefficients, one comes from the symmetry condition $	\dot{e}_{[\beta}^{\-\  \mu }  e_{ \alpha] \mu} = 0$ and reads \cite{Adamo:2017nia} $	 (b^{\uparrow, \downarrow } )^\alpha_{\-\ [\mu} (c^{\uparrow, \downarrow } )_{\alpha \nu ]} = 0$, the other one comes from the conservation of the Wronskian related to the harmonic oscillator equation and links in- and out-states \cite{Garriga:1990dp} $b_{\alpha \mu } ^{\downarrow } \delta^\alpha_\nu =- b_{\alpha \nu}^{\uparrow} \delta^{\alpha}_\mu$. It is worth recalling that the coefficients $b^\updownarrow$ represent the velocity memory effect induced by the passage of the wave.
	Now that the states are well defined we should introduce a scalar product, however the standard definition \cite{birrell_davies_1982} needs a well-defined Cauchy surface $\Sigma$ 
	\[
	\braket{\Psi_2}{\Psi_1} =  - i \int_\Sigma \sqrt{ \mathcal{G}_\Sigma} d \Sigma^\mu 
	\Psi_2 \overleftrightarrow{\partial}_\mu \Psi_1^*
	.\]
	As we have already discussed, it is impossible to define a Cauchy surface $\Sigma$ in this spacetime \cite{Penrose:1965rx}, nevertheless the surfaces of constant $\phi$ are intersected by almost all the geodesics except the ones with constant
	$\phi$ \cite{Gibbons:1975jb}, therefore they can be chosen to construct a foliation of the spacetime. We can thus choose an arbitrary $\phi$ and rewrite the product definition as
	\[
	\braket{\Psi_2}{\Psi_1} =  - i \int_{\phi = const.}  d X^+ d^2 X_\perp 
	\Psi_2 \overleftrightarrow{\partial}_+ \Psi_1^*,
	\]
	where we recall that the absolute value of the Brinkmann determinant is always equal to one due to its Kerr-Schild form. A straightforward calculation shows that the Bogoliubov coefficient encoding particle creation is zero. This corresponds to the amplitude for a positive frequency in-state to develop a negative frequency out-state component:  $\braket{\Phi^\downarrow_1}{\Phi^{\uparrow*}_2} = 0$ \cite{Gibbons:1975jb,Garriga:1990dp,Adamo:2017nia}. It follows that the in- and out-vacua can be identified and no particle is created by
    the background, as in the electromagnetic case. On the other hand, one can show that the in-to-out scattering probability depends on the determinant of $\dot{e}^\uparrow_{ij}$ in the out-region \cite{Garriga:1990dp}
    \[
	\left.
	\abs{\braket{\Phi^\downarrow_1}{\Phi^{\uparrow}_2}}^2 \propto \frac{\delta(p_1^- -p_2^-)}{|\det \dot{e}^\uparrow_{ij}|}\right\vert_{\text{out-region}}.
	\]
	Thus the probability for a scalar field to be scattered from the gravitational wave background is directly correlated to the velocity memory effect. This is what one would expect, because the difference between the initial and final momenta induced by the background corresponds to this effect. The same is true for the classical cross section describing the interaction between a particle and the gravitational wave \cite{Garriga:1990dp}.

	\subsection{S-matrix element}
	In the previous sections we prepared all the tools needed in order investigate the emission of a single photon by an electron (Compton scattering) in a sandwich plane wave spacetime. The amplitude for the process $e(p) \rightarrow e(p')+ \gamma(k')$ in this background is  
    \[
    M = -ie  \int d^4 X  \bar{\Psi}^\downarrow_{p'} \gamma^\mu \Psi^\uparrow_p A^{\downarrow *}_{k',\mu}. \]
    As a first consistency check it is interesting to consider its linear order in $\kappa$ in the monochromatic assumption. This has to coincide with the known results for the inverse graviton photoproduction $g(k) + e(p) \rightarrow e(p')+ \gamma(k')$ in vacuum quantum field theory, where $k^\mu = \omega n^\mu$ is the momentum of the gravitational wave. In this approximation we are free to use the quantum states in Rosen coordinates, in fact the difference between the in- and the out-vacua emerges at the second order. We will thus drop the in and out labels and consider 
    $M = -ie  \int d^4 x  \bar{\Psi}_{p'} \bar{\gamma}^\alpha \Psi_p \bar{A}^{*}_{k',\alpha} + \Ogrande(\kappa^2)$, where the states will be defined below and where the metric determinant does not contribute at this order. Everything we need for this approximation is encoded in the vierbein, which is now expanded as 
    \[
    e_{\alpha \mu } = \eta_{\alpha\mu} + \frac{\kappa}{2} h_{\alpha \mu }
	  \quad , \quad  
	e_{\alpha }^{\-\ \mu } = \delta_\alpha ^\mu   - \frac{\kappa}{2} h_{\alpha }^{\-\ \mu },\]
    where $h_{\alpha \mu }$ is the graviton field. We can now introduce the monochromatic assumption defining $h_{\mu \nu} = \varepsilon_\mu \varepsilon_\nu e^{- i k \cdot x} $, where the graviton polarization tensor has been written as a product of two photon polarization vectors $\varepsilon_{\mu\nu}(k) = \varepsilon_\mu(k) \varepsilon_\nu(k) $ satisfying $\varepsilon\cdot \varepsilon = 0$, $k\cdot \varepsilon = 0$ and $\varepsilon \cdot \varepsilon^* = -1$  \cite{Gross_1968}. This together with the energy-momentum conservation leads to the simple substitution rule 
    \[
    P_\alpha \rightarrow  \frac{1}{2}  e^{-i k \cdot x} \varepsilon \cdot p \varepsilon_\alpha
    \]
    in $\Psi_p$, in $\bar{\Psi}_{p'}$ (with $p\to p'$), and in $\bar{A}^{*}_{k',\alpha} $ (with $p\to k'$). With this in mind we can easily find the states under these assumptions in the form
    \[
	&\Psi_p  \simeq  e^{- i p \cdot x }
    \left[
    1 + 
		e^{- i k \cdot x } \kappa\frac{p\cdot \varepsilon }{4 p\cdot k }\left(
		2 p \cdot \varepsilon 
		+  \slashed{k}\slashed{\varepsilon}	\right)
    \right]u_p + \Ogrande(\kappa^2)
 ,
    \\ &
    \bar{A}^*_{k',\alpha} \simeq    
		e^{i k' \cdot  x } \left[
		\varepsilon^{'*}_\alpha  +
		e^{- i k \cdot x}\kappa \frac{k' \cdot \varepsilon  }{2 k \cdot k'}
		\left(
		\varepsilon^{'*} \cdot \varepsilon k _\alpha - 
		k' \cdot \varepsilon   \varepsilon^{'*}_\alpha 
		- \varepsilon^{'*}\cdot k \varepsilon_\alpha 
		\right)
		\right] + \Ogrande(\kappa^2).
	\]
    Now, one can plug these expressions into the amplitude previously defined. Introducing the notation $M = (2 \pi)^4 \delta^{(4)}(p+k-p'-k')\mathcal{A} + \Ogrande(\kappa^2)$, the following result is found after a proper rearrangement of the various terms 
	\[
	\mathcal{A} = 
	- i e \frac{\kappa}{2} \bar{u}_{p'}&\left[
	\frac{ p' \cdot \varepsilon }{2 p' \cdot k }\left(
	-2 p' \cdot \varepsilon 
	+  \slashed{\varepsilon} \slashed{k}	\right)
	\slashed{\varepsilon}^{'*}
	+  
	\frac{p \cdot \varepsilon}{2 p\cdot k }\slashed{\varepsilon}^{'*}\left(
	2 p \cdot \varepsilon 
	+  \slashed{k}\slashed{\varepsilon}	\right)  
	\right. \\  & \qquad \left.
	+\frac{\varepsilon \cdot k' }{ k \cdot k'}
	\left(
	\varepsilon^{'*} \cdot \varepsilon \slashed{k} - 
	\varepsilon \cdot k'  \slashed{\varepsilon}^{'*}
	- \varepsilon^{'*}\cdot k \slashed{\varepsilon} 
	\right)
	\right]u_p,
	\]
    which agrees with the vacuum QFT result as expected (see e.g. \cite{Choi_1994,Holstein_2006_b}). The first-order expansion of the dressed fermions naturally generates the $s$ and $u$ channels contributions, while the expansion of the photon state reduces to the sum of the $\gamma$-pole and the seagull diagrams \cite{Choi_1994}. The contact term never shows up explicitly in our calculation as a consequence of graviton gauge invariance \cite{Choi:1993_c}: the seagull diagram only provides a momentum-independent contribution which cancels out a term of the same kind in the $\gamma$-pole diagram.
    
    Let us now go back to the original amplitude. In the fully nonlinear case we have to work in the Brinkmann chart in order to have a well defined matrix element. Choosing the gauge $\nabla_\mu A^\mu  = 0 = A^-$ for the photon field, the amplitude gets the following form 
    \[
	M = -ie  \int d^4 X  
	(\Omega^\downarrow)^2 \Omega^\uparrow
	e^{i ( S^\uparrow_p-S^\downarrow_{p'} -S^\downarrow_{k'}  )}
	\bar{u}_{p'}
	\left(
	1 - \frac{\slashed{\Delta}^\downarrow_{p'} \slashed{k}}{2k\cdot p'} 
	\right)
	\left(
	\slashed{\varepsilon}^*
	- \frac{ \varepsilon^{i*} \Delta^\downarrow_{k',i}\slashed{k} }{ k\cdot k'} 
	\right)
	\left(
	1 - \frac{\slashed{k}\slashed{\Delta}^\uparrow_{p} }{2k\cdot p} 
	\right)u_p,
	\]
	The unpolarized squared amplitude averaged over the initial spin reads 
    \[
&	\frac{1}{2}\sum_{s, s',\varepsilon'} M^{\dagger } M   =  \frac{e^2}{2} 
	\int d^4 X \mathfrak{I}(X)  \int d^4 X  '
	\mathfrak{I}^*(X')\times \\&
	\quad\times \text{Tr}\left[
		(\slashed{p}' + m )
		\left(
		1 - \frac{ \slashed{\Delta}^\downarrow_{p'}(X)\slashed{k}}{2 k \cdot p' }
		\right) 
		\left(
		\bar{\gamma}^\alpha
		- \frac{  \Delta^{\downarrow,\alpha}_{k'}(X)\slashed{k} }{ k\cdot k'} 
		\right)
		\left(
		1 - \frac{\slashed{k}\slashed{\Delta}^\uparrow_p(X)}{2 k \cdot p }
		\right)
		\right.
		\\ & \quad\times 
		\left.
		(\slashed{p} + m )
		\left(
		1 - \frac{\slashed{\Delta}_p^{\uparrow}(X')\slashed{k}}{2 k \cdot p }
		\right) 
		\left(
		\bar{\gamma}^\beta
		- \frac{  \Delta^{\downarrow,\beta}_{k'}(X ')\slashed{k} }{ k\cdot k'} 
		\right)
		\left(
		1 - \frac{\slashed{k}\slashed{\Delta}_{p'}^{\downarrow}(X')}{2 k \cdot p' }
		\right) 
	\right]  
	\left(
	- \eta_{\alpha \beta} 
	+ \frac{k_\alpha k'_\beta + k'_\alpha k_\beta}{ k\cdot k' }
	\right),
	\]
    where the prefactor is defined as $\mathfrak{I} = (\Omega^\downarrow)^2 \Omega^\uparrow
	e^{i ( S^\uparrow_p-S^\downarrow_{p'} -S^\downarrow_{k'}  )}$. 
     Once the trace is computed this expression reduces to the following form 
	\[
	\frac{1}{2}\sum_{s, s',\varepsilon'} M^{\dagger } M   = & \-\  \frac{e^2}{2} 
	\int d^4 X \mathfrak{I}(X)  \int d^4 X  '
	\mathfrak{I}^*(X') \\&
	\times 
	\bigg\{
	\frac{k \cdot p'}{k \cdot k'}\left(
	\Pi^{\uparrow i}_{p}(X) \Pi^{\downarrow  }_{k',i}(X')
	+ 
	\Pi^{\downarrow i}_{k'}(X) \Pi^{\uparrow  }_{p,i}(X')
	+ 2 p^+ k'^- +2 p^- k'^+
	\right)
	 \\&  +
	\frac{k \cdot p}{k \cdot k'}\left(
	\Pi^{\downarrow i}_{p'}(X) \Pi^{\downarrow  }_{k',i}(X')
	+
	\Pi^{\downarrow i}_{k'}(X) \Pi^{\uparrow }_{p',i}(X')
		+ 2 p'^+ k'^- +2 p'^- k'^+
	\right)
	\\&  -
	 \frac{k\cdot p '}{ k\cdot p}	\left(
	 \Pi^{\uparrow i}_{p}(X)\Pi^{\uparrow }_{p,i}(X')
	 + 2 p^+ p^-
	 \right)
	- 
	\frac{k\cdot p }{ k\cdot p'}
	\left(
	\Pi^{\downarrow i}_{p'}(X)\Pi^{\downarrow }_{p',i}(X')
	+ 2 p'^+ p'^-
	\right)
	\\&  -
	2\frac{k \cdot p k\cdot p '}{ (k\cdot k')^2}
	\left(
	\Pi^{\downarrow i}_{k'}(X)\Pi^{\downarrow }_{k',i}(X')
	+ 2 k'^+ k'^-
	\right)
	+ \frac{ m^2  (k\cdot k')^2  }{k \cdot p  k 
		\cdot p'}
	\bigg\}.
	\]
	It is interesting to observe that the terms in the amplitude quadratic in the gravitational field disappear in the unpolarized sum. For this reason, the integrals in the three coordinates $X^+,X^i$ are easily carried out by observing that the prefactor is Gaussian in $X^i$. Namely 
	\[
	\int & d^4 X \mathfrak{I}(X)  = 
	-i (2 \pi)^{2} \delta(p^- -p'^- -k'^-)  
    \int d\phi
	\frac{(\Omega^\downarrow)^2 \Omega^\uparrow }{ p^-\sqrt{C } }
 e^{-i(p^+ -p'^+ - k'^+) \phi}
    \\ & 
	\times
	e^{  \frac{i}{2 p^- }
		B^i(C^{-1})_{i j} B^j	
	- i \frac{\kappa}{p^-} \int^{\phi}_{- \infty} d \tilde{\phi}
    \left(
    p_\alpha P^\alpha - \frac{\kappa}{2} P_\alpha P^\alpha
    \right)
	+  i \frac{\kappa}{p'^-} \int^{\phi}_{ \infty} d \tilde{\phi}
    \left(
    p'_\alpha P'^\alpha - \frac{\kappa}{2} P'_\alpha P'^\alpha
    \right)
    +  i \frac{\kappa}{k'^-} \int^{\phi}_{ \infty} d \tilde{\phi}
    \left(
    k'_\alpha K'^\alpha - \frac{\kappa}{2} K'_\alpha K'^\alpha
    \right)}
	 ,
	\]
	where $C_{ij} = \sigma^\uparrow _{ij}- \sigma^\downarrow_{ij}$, $C = \det C_{ij}$ and $B_i = e^{\uparrow j}_i p_j -e^{\downarrow j}_i( p'_j  + k'_j) $. By setting $
	\int d^4 X \mathfrak{I}(X)  =
	i (2 \pi)^{2} \delta(p^- -p'^- -k'^-) 
	\int d \phi \tilde{\mathfrak{I}}(\phi)$, the only other integral needed is the one with a prefactor linear in the transverse coordinate $X^i$, namely 
	\[
	\int d^4 X \mathfrak{I}(X) X^i =
	i (2 \pi)^{2} \delta(p^- -p'^- -k'^-) 
	\int d \phi \tilde{\mathfrak{I}}(\phi)
	\frac{(C^{-1})^i_{\-\ j} B^j}{p^-}	.
	\]
	We can finally write down the partially integrated transverse momentum as
	\[
	\int d^4 X \mathfrak{I}(X) \Pi^{\uparrow i}_p(X) = 
	-i (2 \pi)^{2} \delta(p^- -p'^- -k'^-) 
	\int d \phi \tilde{\mathfrak{I}}(\phi)
	\left(
	e^{\uparrow ij}p_j -  \sigma^{\uparrow i}_j
		(C^{-1})^j_{\-\ k} B^k	
	\right),
	\]
    with this substitution rule the squared amplitude can be written as an integral in $\phi,\phi'$ only. A further simplification can be achieved exploiting the identity  $
	C^{-1}_{ij} = \frac{1}{C}(  C_k^{\-\ k}\eta_{ij} - C_{ij}        )
	$ for rank-two matrices. Having derived the full nonlinear squared amplitude it would be natural to consider some physically interesting examples, however one runs very soon into the mathematical complexity of the problem.
    In particular, we recall that once the Brinkmann profile is chosen one has to solve the differential equation $\ddot{e}_{ij} = H_{ik}e^k_{\-\ j}$, in fact the knowledge of the in and out vierbein is a key element. There are very few profiles $H_{ij}$ for which an analytical solution of this equation is available. These include for example impulsive waves $H_{ij} = \delta(\phi) d_{ij}$ where $d_{ij}$ is a constant matrix, these solutions are relevant as they represent the gravitational field produced by a massless particle moving at the speed of light \cite{Aichelburg:1970dh, Penrose:1972xrn, Dray:1984ha}. Other possibilities are profiles with conformal symmetry of the form $H_{ij} = a(\phi^2 + b^2)^{-1}\epsilon_{ij}$ studied in \cite{Andrzejewski:2018pwq}. However physically interesting scenarios for the process considered here would require proper approximations and numerical evaluations. Also the concept of formation length, of great importance in electromagnetism, could be worth to be studied in this context. In fact, despite the non-locality of the interaction and the clear implications of a curved spacetime background, the high amount of symmetries of the problem could lead to some interesting developments. A detailed study of this and similar processes in the aforementioned directions will be left for future works.

    \section{Conclusions and outlook}
	We examined the one-to-one map connecting the dynamics of a charged particle in an electromagnetic plane wave and a particle in a nonlinear gravitational plane wave. This map enables to translate interesting results known in electromagnetism to the context of plane wave spacetimes. The dynamics in an electromagnetic plane wave is completely determined by a specific Lorentz-like transformation $\Lambda^\alpha_{\-\ \beta}$, which provides a way to transport the initial momentum of a free particle through the wave such that $\pi^\alpha =\Lambda^\alpha_{\-\ \beta} p^\beta $. This transformation has a number of interesting properties. Besides belonging to the Lorentz group, it operates on constant-$\phi$ hypersurfaces and it acts as a gauge transformation on the background wave. For these reasons its proper generalization as a quantum operator $\mathbf{\Lambda}$ turns out to be a key element to construct quantum states in an electromagnetic plane wave background. Exploiting the aforementioned analogy between electromagnetic waves in flat spacetime and gravitational nonlinear waves we showed that the curved spacetime generalization of $\mathbf{\Lambda}$ can be used to find quantum states in gravitational plane waves as well. In particular we showed how scalar, spinor, and vector fields can be naturally mapped from one context to the other. 
    Finally, in the last part of the paper we applied the found states for spin 1/2 and spin 1 particles to consider Compton scattering in a nonlinear sandwich gravitational wave. Here we provided the spin- and polarization-summed squared amplitude for the process, exact in the gravitational plane wave. A detailed study of the Compton cross section for physically interesting plane wave gravitational backgrounds is left for a future work. 
    Other processes could be naturally investigated as well in the future, an interesting example being the graviton emission by a moving particle. This could provide insights about its relation with photon emission, a relation which has been studied in vacuum QFT \cite{Choi_1994,Holstein_2006_a} and in the presence of a background electromagnetic plane wave within strong-field QED \cite{Audagnotto:2022, Grats_1975}. 
	
	\section*{Acknowledgments}
	The authors would like to thank C. H. Keitel and L. Tamburino Ventimiglia di Monteforte for reading the paper and for the valuable comments.

    \newpage
    \appendix
	\section{Metrics handbook} \label{App. metrics}
 In this appendix, we collect a few known but useful formulas about the Brinkmann and the Rosen metrics. We refer to the main text for the notation.
	\begin{itemize}
		\item
		\textbf{Brinkmann} 
		\begin{itemize}
			\item The metric: $ds^2 = \eta_{\mu\nu}dX^\mu dX^\nu + H d\phi^2$
			\item Christoffel symbols: $\Gamma^\lambda_{\mu\nu} = 
			n^\lambda \left(
			\frac{1}{2}\dot{H} n_\mu n_\nu + 2 H_{(\mu }  n_{ \nu)}
			\right)
			- H^\lambda n_\mu n_\nu$
			\item Riemann tensor: $R_{\rho \mu \nu \delta} = 
            \left[
            H_{\rho \nu} n_\mu n_\delta 	-(\rho \leftrightarrow \mu) \right] - (\nu \leftrightarrow \delta)$
			\item Ricci tensor: $R_{\mu \delta} = H^i_{\-\ i} n_\mu n_\delta $
			\item Vierbein: $E_{\alpha \mu} = \eta_{\alpha\mu} + \frac{H}{2}n_\alpha n_\mu$
			\item Spin connection: $\omega_{\mu \alpha \beta }=  2 n_\mu H_{[\beta}  n_{\alpha]} $
		\end{itemize} 
		where $H = H_{ij}dX^idX^j$, $H_\mu = H_{\mu i} X^i$.
		\item
		\textbf{Rosen}
		\begin{itemize}
			\item The metric: $ds^2 = 2dx^+d\phi +  \gamma_{ij}dx^idx^j $
			\item Christoffel symbols: $\Gamma^\lambda_{\mu\nu}  = 
			\frac{1}{2}g^{\lambda \rho }\left(
			n_\mu \dot{\gamma}_{\rho \nu } +	n_\nu \dot{\gamma}_{\rho \mu } 
			-	n_\rho \dot{\gamma}_{\mu \nu }
			\right) $
			\item Riemann tensor: $R_{\rho \mu \nu \delta} = 
			\left[\left(\frac{1}{4}\gamma_{\rho \sigma}\dot{\gamma}^{\sigma \lambda}\dot{\gamma}_{\lambda \nu } +
			\frac{1}{2}\ddot{\gamma}_{\rho \nu }\right) n_\mu n_\delta
			-(\rho \leftrightarrow \mu) 
               \right]
            - (\nu \leftrightarrow \delta)$
			\item Ricci tensor: $R_{\mu \delta} = 
			\left(\frac{1}{4}\dot{\gamma}^{\rho \lambda}\dot{\gamma}_{\lambda \rho } +
			\frac{1}{2}\gamma^{\rho \lambda} \ddot{\gamma}_{\lambda \rho }\right) n_\mu n_\delta   $
			\item Vierbein: $e_{\alpha\mu}$
			\item Spin connection: $\omega_{\mu \alpha \beta }= n_{[\alpha}\dot{e}_{\beta] \mu}$	
            \item 
            Spin connection: $\omega_{\mu \alpha \beta }= - 2 n_{[\alpha}\dot{e}_{\beta] \mu}$	
		\end{itemize}  
	\end{itemize}

	\newpage


%

\end{document}